# Influence of oxygen reduction on the structural and electronic properties of electron-doped $Sr_{1-x}La_xCuO_2$ thin films

Z. Z. Li[a], V. Jovanovic[a] , *H. Raffy[a], S. Megtert[b]

[a]Laboratoire de Physique des Solides, Bât. 510, UMR 8502-CNRS, Université Paris-Sud, F-91405 Orsay Cedex, France

[b]Unité mixte de Physique, UMR137 CNRS/Thalès, F-91767 Palaiseau Cedex, France

**Abstract**

Single phase, c-axis oriented, e-doped, $Sr_{1-x}La_xCuO_2$ thin films were epitaxially grown on KTaO$_3$ and DyScO$_3$ substrates by reactive rf sputtering. As-grown films being insulating due to oxygen excess, oxygen reduction is necessary to observe superconductivity. Two different procedures were employed to reach superconductivity. On one hand an in-situ reduction process was conducted on a series of films deposited on both types of substrates. On the other hand, an ex-situ reduction procedure was performed sequentially on a single film deposited on DyScO$_3$. The study of the influence of oxygen reduction on the structural and electronic properties of the thin films is presented and discussed.

*\*)* Corresponding author. Address : Laboratoire de Physique des Solides, Bât. 510, Université Paris-Sud, F-91405 Orsay Cedex, France . Tel :.33 1 6915 5338; fax : 33 1 69 15 6086.  e-mail address : raffy@lps.u-psud.fr

# 1. Introduction

The electron-doped infinite layer (IL) compound $Sr_{1-x}Ln_xCuO_2$, (Ln=La, Nd and Pr ; $T_{cmax}$= 43K for Ln=La) [1,2] has the simplest crystallographic structure among all of the superconducting cuprates. It is only constituted of stacked $CuO_2$ planes, composed of $Cu^{2+}$ ions coordinated by fourfold oxygen, responsible of superconductivity, with Sr, Ln atoms as inter-plane separators. In the ideal structure [3] no Cu apical oxygen is present and there is no charge reservoir. The electron charge carriers are supplied by $Ln^{3+}/Sr^{2+}$ substitution. The parent compound of this phase is $SrCuO_2$ (x=0). Prepared at ambient pressure, $SrCuO_2$ has an orthorhombic structure and contains double chains $Cu_2O_2$ of square planar groups sharing edges joined through one rock-salt type SrO layer [4,5]. To obtain the electron-doped superconductor, one first needs to obtain the compound with $CuO_2$ planes and with increased Cu-O-Cu bond length, lattice constant $a$, in order to favor the electron carrier doping of the $CuO_2$ planes. The substitution of $Sr^{2+}$ by $Ln^{3+}$ expands the Cu-O bond, and therefore the lattice constant $a$ increases. The second condition to get this material superconducting is to remove excess apical oxygen without removing oxygen in the $CuO_2$ planes. The formation of oxygen vacancies in $CuO_2$ layers, resulting in the so-called "long $c$" phase ($c$~3.62Å) [6] is the second cause for decreasing or suppressing superconductivity in the $Sr_{1-x}Ln_xCuO_2$ system. Unfortunately these conditions are not easy to meet. The synthesis of IL bulk ceramic material is difficult, requiring very high pressure (2-6 GPa) [7,8]. No single crystal has been prepared so far. However, these conditions can be satisfied using appropriate substrates to take benefit of the epitaxy effect where interfacial interactions tend to force the deposited thin film in-plane lattice parameters to match the substrate ones. Doing so, IL $Sr_{1-x}La_xCuO_2$ thin films offer the unique opportunity to obtain oriented samples needed for physical studies. However very few groups have prepared these films. Superconducting $Sr_{1-x}La_xCuO_2$ thin films have been deposited by MBE [9] and by laser ablation [10]. Our films are deposited by a sputtering technique, that we also employ for hole-doped cuprate thin films [11]. Unfortunately the production of an imperfect distribution of oxygen in the IL compound during the film deposition is unavoidable. In our process as in the MBE process of ref.9, excess apical oxygen atoms are present between $CuO_2$ planes, while in the process reported in ref. 10, there are oxygen vacancies in the $CuO_2$ planes. In our case as in ref. 9, a reducing oxygen process to remove apical oxygen is indispensable to get superconducting films. This reducing procedure generally takes place in-situ during the cooling process, after the deposition at high temperature. We will show that it is also possible to reduce the oxygen content of one $Sr_{1-x}La_xCuO_2$ film by applying an ex-situ procedure, which offers the advantage to study the evolution *vs* doping of a single film as previously reported for hole-doped BiSrCaCuO thin films[13].

Our paper is organized as follows. We first describe our techniques of synthesis and characterization of $Sr_{1-x}La_xCuO_2$ thin films. We report how we explore the influence of oxygen reduction on the structural and electronic properties of electron-doped $Sr_{1-x}La_xCuO_2$ (x=0.12) thin films deposited on $KTaO_3$ and $DyScO_3$ substrates respectively, by using two different post-deposition reducing oxygen procedures, either in-situ or ex-situ. We describe, in this letter, three situations. The first two situations are related to samples for which we examine the influence of the temperature $T_A$ of an *in-situ* oxygen reduction period. The third situation describes the evolution of one insulating (as-prepared) $Sr_{0.88}La_{0.12}CuO_2$ film on $DyScO_3$ substrate, which was repeatedly ex-situ vacuum annealed.

## 2. Experimental

### A. Synthesis

Epitaxial *c*-axis oriented thin films, 600-650Å thick, were prepared by on-axis, single target, *rf* magnetron sputtering on heated (690°C) single crystal substrates. The choice of the substrate plays a very important and determinant role to obtain electron-doped IL superconducting thin films. The traditional substrate $SrTiO_3$ (STO), widely used for hole/electron-doped high-$T_c$ superconducting cuprate deposits, is unsuitable for this compound due to its lattice parameter (a = 3.905 Å) shorter than that of the IL compound (*a* ~ 3.95 Å for bulk samples with x = 0.12)[2]. So far, good candidate substrates are (100) $KTaO_3$ (KTO, a = 3.989 Å) and (110) $DyScO_3$ (DSO, $a_0$ = 3.9435 Å) or (100) $SrTiO_3$ with (100) $BaTiO_3$ or (001) $Pr_2CuO_4$ buffer layer [10,14]. Properties of the substrate surface are very important parameters as well. They affect the quality, reproducibility and stability of the thin film. In the presented paper, films were deposited on (100) KTO and (110) DSO substrates. KTO substrates were cleaned with acetone and methanol and further heated in air at 650° C for several hours. After heat treatment, substrates are immediately stuck on the substrate holder with Ag epoxy glue, and the holder set into the vacuum system. For DSO substrates, after being treated at 950°C in air for 24 hours, they were etched using a solution of 0.1% $HNO_3$ in methanol as described in ref.9. During the deposition, the sputtering gas was a mixture of $Ar/O_2$ (4 to 1) and the total pressure was fixed at 200 mTorr. The distance between the target and the substrate was 40 mm, the rate of deposition was about 0.20 Å/s. In this paper, we study the influence of an in-situ oxygen reduction procedure which takes place during the cooling process and consisted to submit the samples to an annealing stage under vacuum (about $2 \times 10^{-6}$ Torr) at given temperature $T_A$ (450°C-625°C) and given time extent $t_A$ (10mn in general). The influence of $T_A$ will be detailed in the description of the results. Otherwise, the samples not reduced in-situ or devoted to ex-situ post-annealing experiments were directly cooled under gas process (GP) down to room temperature.

*B. Techniques of characterization*

The structural properties of the deposited films were studied by X-Ray diffraction (XRD) : $\theta$–$2\theta$ scans which allow to identify the phases present in the deposit and to determine $c$ axis parameter value, $\Phi$-scans to check epitaxial growth and $\omega$-scans to control the mosaïcity. The morphology and roughness of surface samples were also observed by atomic force microscopy and interferential optical microscope. The film composition was determined by inductive coupled plasma spectroscopy (ICP). The thickness of the films was measured by Rutherford Backscattering Spectroscopy (RBS) and X-ray reflectivity. A standard *dc* four probe method, with sputtered gold pads, was used for resistance *vs* temperature measurements, $R(T)$, of the sample, using a home made dipping stick inside a helium dewar (4.2 K$\leq$ T $\leq$ 300 K). The results were compared with magnetization and *ac* susceptibility measurements performed with a SQUID magnetometer. Some films were patterned by electron lithography and chemical etching to study their transport properties, Hall effect and magneto-resistance, which will be presented in a subsequent paper.

### 3. Results and discussion

*A. In-situ oxygen reduction*

1.  $Sr_{1-x}La_xCuO_2$ thin films on $KTaO_3$ substrates.

A series of films (x = 0.12) were deposited under the same sputtering conditions, as indicated above, on $KTaO_3$ substrates. After a stage at 670°C for 30 minutes under gas process, the samples were submitted to an in-situ annealing period at given temperature $T_A$ (450°C-625°C) under a reduced oxygen pressure (about $2x10^{-6}$ Torr). The annealing period lasts 10 mn. Each film of the series is characterized by the temperature $T_A$ at which it was annealed. Figure 1 displays the evolution of the temperature dependence of the resistance $R(T)$ of these films with increasing $T_A$. Without this reduction stage (GP curve) or for too low $T_A$ ($T_A$< 500 °C), $R(T)$ exhibits high values with an insulating behaviour. By increasing the annealing temperature $T_A$, the resistance decreases drastically and its behaviour is changed. For $T_A$ above 500°C a metallic decrease of $R(T)$ is first observed at high temperature followed by a non metallic logarithmic (*Ln T*) increase, foreseeing a superconducting transition. The temperature $T_{min}$ (see vertical arrows) where $R(T)$ is minimum, below which the low temperature upturn in resistance takes place, decreases with increasing $T_A$ whereas the superconducting transition temperature $T_c$ increases.

For $T_A \leq 625$ °C, the films are single-phase and *c*-axis oriented. A typical $\theta-2\theta$, Cu-Kα, XRD pattern is displayed in Figure 2 (with, in insert, a phi-scan showing that the film is epitaxially grown). Only (*00l*) Bragg reflections of the IL phase and of the substrates were detected, as evidenced with the logarithmic *y*-scale to look at tiny details of the background (at this scale one can even see sharp cuts on the left side of the tails of the intense substrate reflections that are the signature of copper absorption edge of the used anode X-ray tube emission). No parasitic phase is present, even in the background. The lattice constant $c$ was determined from ($002$) Bragg reflection angular position. As for the resistance, the value of lattice parameter $c$ is dependent on the cooling conditions. It decreases monotonically with increasing $T_A$, as shown by the shift to higher angles of the ($002$) reflection (see fig. 3 and insert) and superconductivity appears when $c$ is less than 3.41 Å. This indicates that by this reducing oxygen procedure, apical oxygen atoms are removed and, as a consequence, parameter $c$ decreases.

In order to reduce further the oxygen partial pressure during the cooling process, oxygen gas supply was turned off at 690°C, when the deposition was stopped, and the samples cooled in argon (instead of process gas) down to the annealing stage. Then argon gas supply was turned off to anneal the samples under vacuum at $T_A$. Figure 4 (a) shows $R(T)$ curves of a series of such films which underwent a vacuum annealing procedure at various $T_A$ for 10 mn. It appears that superconductivity is reached at a lower $T_A$ value than that it was required in the previous process. For instance, a complete superconducting transition at 13K is obtained with $T_A = 550$ °C while only the transition onset was visible at 4.2K in the first case (see Fig. 1). It is to be noted that $T_c$ increases while $c$ decreases with increasing $T_A$ (see insert in fig.4 (a)). The curve with $T_c$ (R=0) = 20 K is quite similar to the one shown in ref.9, in which the sample was treated under a pressure of $1.10^{-5}$ mbar, comparable to the value of the pressure in our reduction stage ($2 \times 10^{-6}$ Torr). This corroborates the fact that reducing oxygen pressure is one important key parameter to get higher $T_c$.

Finally, Figure 4 (b) displays, for a similar film on KTO with $T_c$ (R = 0) ~ 20 K, both the resistivity curve $R(T)$ and the magnetic transition $\chi(T)$ measured by *ac* susceptibility. The latter confirms the quality of the superconducting state in such films.

## 2. $Sr_{1-x}La_xCuO_2$ thin films on $DyScO_3$ substrates

Figure 5 shows the temperature dependence of the resistance for films (x = 0.12) which were deposited on DSO substrates under the same conditions as those for films on KTO, and underwent different conditions during in-situ post-annealing reducing oxygen process. They were cooled down to a temperature $T_A$ under *Ar* atmosphere and the samples kept at $T_A$ for 10 mn. under vacuum (at about $2x10^{-6}$ *Torr)*. Compared to films on KTO substrates, it should be realized that, in order to avoid the presence of the "long-$c$" phase, the $T_A$ values have to be much lower than those used for the films on KTO substrate. For instance, while we do not see the presence of the "long $c$" phase in films on KTO even for $T_A$ = 625 °C, this phase already appears in the film on DSO above $T_A$ = 570 °C. Also it is seen (fig.6(a)) that the presence of the "long $c$" phase, estimated from the ratio, $I_{lc}/I_{IL}$, of the intensity of (002) Bragg reflection for each phase ("long-$c$" and IL phases respectively), strongly increases for $T_A$ > 575 °C (annealing time 10 mn.). By varying the duration of annealing (1mn. to 30 mn.), it is also seen that the ratio $I_{lc}/I_{IL}$ decreases by reducing this duration. But for high $T_A$ values ( > 590 °C) and very short anneal time length (1mn.) the value of $T_c$ decreases (fig.6(b)) and $c$ increases (fig. 6 (a)) probably due to the fact that a short annealing time is not sufficient to remove apical oxygen. Examples of XRD patterns and their simulation of films on DSO substrate are shown in Fig.7 (see also Fig.1 in ref.12). We can see that for films with or without "long $c$" phase on DSO substrate, intensity oscillations around the main reflections are clearly observed, meaning that layers are very flat, smooth and with well defined interface with the substrate. Such oscillations were less visible for films deposited onto KTO substrates. This could be due to the small mismatch between the film and the DSO substrate lattice parameters, (very close to that of the bulk SLCO material) resulting in the best condition for epitaxy as can be seen for example with a rocking curve scan (insert of Fig.5). The full width at half-maximum (FWHM) of films on DSO is in the range 0.04-0.07°, a value comparable to that of the substrate. This indicates low interfacial stress and induced deformations between film and substrate.

*B. Ex-situ oxygen reducing process*

1. Evolution of a $Sr_{1-x}La_xCuO_2$ film on DSO substrate

In this paragraph, it is shown how an ex-situ annealing process applied to a $Sr_{1-x}La_xCuO_2$ film (x = 0.12) deposited on a DSO substrate allowed us to change the electrical film properties from an insulator in the as-deposited state to metallic and superconducting phases. For this purpose, one sample was deposited onto a DSO substrate, cooled down under gas process without any in-situ reducing oxygen stage. Then the sample was submitted to a sequence of low temperature annealing treatments in a vacuum tube furnace ($1.5$-$2x10^{-6}$ *Torr*). It was successively annealed in 23 steps, first one at 265°C (step1), then at 300°C (step 2) for 15min. and for the following treatments it was kept at 320°C with annealing duration ranging from 15 to 45 min. (steps 3 to 23). After each treatment, *R(T)* and XRD measurements were carried out on the sample. Figure 8 displays the curves showing the evolution of the temperature dependence of the resistance of the film after the different annealing treatments. In the as-prepared state, the resistance, $R_{in}$ $(T)$, of this sample increases strongly with decreasing temperature. Figure 9 (a) displays the data plotted as $Ln$ ( $R_{in}$ ) vs $1/T^{\alpha}$. It appears that they can be described by a function of the form $R= R_0 \exp(T_0/T)^{\alpha}$ characteristic of hopping behaviour of the conduction as described in [15]. The value of $\alpha$ giving a straight line over the largest temperature interval ( 25K<T<230K) is equal to $\alpha = (0.18\pm0.02)$. A hopping behaviour of the conduction is also observed for a film on KTO cooled down under gas process (see Fig.1) as well as for a strongly deoxidized hole-doped $Bi_2Sr_{1.6}La_{0.4}CuO_y$ thin film [13]. The first heat treatment produces a dramatic decrease of the resistance at 300 K, $R_{300K}$, which will continue to monotonically decrease after each treatment. After the third treatment, the low temperature behaviour of the resistance changes from insulating-like (see $R_2(T)$) to superconducting (see $R_3(T)$) and this transition from an insulating to a superconducting state (SIT) is foreseen by a $lnT$ behaviour of the resistance (10 K< T < 200 K) in the state 2 (see Fig. 9 (b) ). A similar $lnT$ behaviour of $R(T)$ was also observed at the SIT transition of a superconducting $Bi_2Sr_{0.6}La_{0.4}CuO_y$ thin film, turned insulating by sequential de-oxidation[13] and also under 60 *Teslas* in Bi(La)-2201 single crystals[16] close to the SIT transition. Starting from step 3, $T_c$ increases with increasing treatment index $i$ for 3 $\leq i<$ 13. Then from step 13, the critical temperature appears to saturate (Fig. 8). At the same time, it also appears that the resistance variation ($R_{i-1}$- $R_i$) becomes $T$ independent, which possibly indicates that only the residual resistance ($R_0$), i.e. the disorder, is decreased by the annealing treatments (13 $\leq i \leq$ 23). This variation tends to zero with the treatment index number. We can therefore conclude that the influence of the annealing treatment in reducing oxygen content is very important at the beginning of the series of thermal treatments, and is less and less efficient after. This conclusion is in agreement with XRD measurements. In Figure 10, are plotted the values of the lattice parameter $c$, after each treatment, versus the corresponding conductance at 300K, along with the onset $T_{on}$ and offset $T_c$ $(R=0)$ critical temperatures. The lattice parameter $c$ decreases very quickly at the beginning and then saturates. Notice that its limit value remains higher than that of films on KTO, where it is less than 3.41 Å. The maximum value of $T_{on}$ is comparable to that observed in the in-situ process. As in the case of the in-situ treatment described above, a limiting factor for increasing $T_c$ is possibly the value of the pressure, not low enough, during the thermal treatment, which is comparable in both cases.

## 4. Conclusion

In the ideal IL $Sr_{1-x}La_xCuO_2$ compound made of stacked planes of $Cu^{2+}$ ions coordinated by fourfold oxygen, and inter-plane (Sr, La) separating atoms, no apical oxygen is present. When present, apical oxygen pushes strongly the compound towards the under-doped region of the phase diagram. Indeed, neutron measurements [17] in another electron-doped compound, $Nd_2CuO_4$, show that there is an average apical oxygen reduction of about 0.06 per unit formula between an oxygenated and a reduced crystal. The main purpose of this paper was to explore the influence of the reducing oxygen process on the structural and electronic properties of electron-doped $Sr_{1-x}La_xCuO_2$ thin films. The reducing oxygen procedure reported here was an heat treatment of the films in low oxygen environment either by an in-situ or an ex-situ method. As-grown films are insulating when cooled in gas process. The in-situ reduction treatment at temperature $T_A$ was applied both on single phase films on KTO ($T_A$ < 620 °C) and on DSO ($T_A$ < 580 °C) substrates. It allowed us to observe the transition from low temperature variable range hopping behaviour of the resistance to $lnT$ behaviour, foreseeing the superconducting transition. The maximum $T_c(R=0)$ reported in this study, equal to 20K (since then increased to 26K), was obtained by an in-situ reduction treatment of films deposited on KTO substrates. It is comparable to the maximum $T_c$ value of the other e-doped family, $Ln_{2-x}Ce_xCuO_4$ (Ln = Pr, Nd ). The main reason why the obtained $T_c$ were lower than the value of the optimal bulk compound appears to be due to the reduction oxygen pressure which is not low enough in our study. As with films grown by MBE [9], where $T_c$ value can be close to the bulk one, it may be possible to improve $T_c$ by reducing the value of the $O_2$ pressure (being actually about 100 times higher compared to the MBE system) during the annealing process. In the case of films on DSO substrate, another reason leading to limited $T_c$ values is the fact that as $T_A$ has to be restricted to low values in order to avoid the presence of "long $c$" phase, oxygen reduction is less effective. Besides it was also shown in ref. 9 that in order to obtain single IL phase (without "long-$c$" phase) films, the upper limit value of the La content x decreases with increasing substrate lattice parameter and one must have : x ≤ 0.08 on STO (a = 3.905Å), x ≤ 0.12 on DSO ($a_0$ = 3.9435 Å) and for x ≤ 0.14 KTO (a = 3.989 Å) respectively. In the same way, to get single IL phase on DSO substrate, a lower (compared to the film on KTO) $T_A$ is necessary. Indeed, we have obtained single IL on this substrate with x = 0.15 by decreasing $T_A$ values (not shown in this paper). Another important result is the fact that our films on KTO substrates are rather stable over at least one year, unlike those mentioned in ref 9 and ref 10. The films on KTO substrates give us the opportunity to study the transport and magnetic properties of the IL compound. The results about the transport properties of IL films will be published elsewhere [18].

## Acknowledgments

V. Jovanovic, on leave from the Institute of Physics of Belgrade, acknowledges support from the E.C., contract n° MEST-CT-2004-514307. We thank L. Fruchter and  F. Bouquet for fruitful discussions. We thank Clarisse Mariet, Laboratoire Pierre Sue, Saclay, for ICP analysis and François Lalue, CSNSM, Orsay,  for RBS analysis.

## References

[1]     M. G. Smith, A.  Manthiram, J.  Zhou, J. B. Goodenough, and J. T. Markert,  Nature (London) **351**, (1991) 549.

[2]     G. Er, Y. Miyamoto, F. Kanamaru, and S. Kikkawa,  Physica C **181**, (1991) 206.

[3]     J. D. Jorgensen, P. G. Radaelli, D. G. Hinks, J. L. Wagner, S. Kikkawa, G. Er, and  F. Kanamaru,  Phys. Rev. B **47**, (1993) 14654.

[4]      Chr. L. Teske and Hk. Muller-Buschbaum, Z. Anorg. Allg. Chem. **379**, (1970)  234.

[5]       M. T. Gambardella, B. Domengès, and B. Raveau, Mat. Res. Bull. **27**, (1992)  629.

[6]    B. Mercey, A. Gupta, M. Hervieu, and B. Raveau,  J. Solid State Chem. **116**, (1995)  37.

[7]      C. U. Jung, J. Y. Kim, S. M. Lee, Y. Yao, S. Y. Lee, and D. H. Ha,  Physica C  **364-365**, (2001) 225.

[8]     A. Podlesnyak, A. Mirmelstein, V. Bobrovskii, V. Voronin, A. Karkin, I. Zhdakhin,

         B. Goshchitskii, E. Midberg, V. Zubkov, T. D'yachkova, E. Khlybov, J. Y. Genoud,

         S. Rosenkranz, F. Fauth, W. Henggeler, and  A. Furrer, Physica C **258**, (1996) 159.

[9]     S. Karimoto, K. Ueda, M. Naito, and T. Imai,  Appl. Phys. Lett. **79**, (2001) 2767;

         **84**, (2004) 2136; Physica C **378-381**, (2002) 127.

 [10]    V. Leca, D. H.A. Blank, G. Rijnders, S. Bals, and G. van Tendeloo,  Appl. Phys.

         Lett. **89**, (2006) 092504; V. Leca, Ph.D. thesis, University of Twente, (2003).

[11]    Z. Z. Li, H. Rifi, A. Vaurès, S. Megtert, and H. Raffy,  Physica C **206**, (1993) 367.

 [12]     Zhi Zhong Li, Iulian Matei, and Hélène Raffy,  Physica C **460-462**, (2007) 452.

[13]    Z. Konstantinovic, Z. Z. Li, and H. Raffy,  Physica C **351**, (2001) 163.

[14]     J. T. Markert , T. C. Messina, B. Dam, J. Huijbregste, J.H. Rector, and R.

         Griessen, *Proceedings* SPIE **4058**, (2000) 141.

[15]     C. Quitmann, D. Andrich, C. Jarchow, M. Fleuster, B. Beschoten, G.

         Guntherodt, V.V. Moschalkov, G. Mante, R. Manzke, Phys. Rev. B **46**, (1992) 11

         813.

[16]     S. Ono, Y. Ando, T. Murayama, F.F. Balakirev, J. B. Betts, and G. S. Boebinger,

          Phys. Rev. Lett. **85**, (2000) 638.

[17]    P. G. Radaelli, J. D. Jorgensen, A. J. Shultz, J. L. Peng, and R. L. Greene,  Phys. Rev.

         B **49**, (1994) 15322.

[18]     V. Jovanovic, Z. Z. Li, F. Bouquet, L. Fruchter,  and H. Raffy, in Proceedings of the

         International Conference on Low Temperature Physics, LT25, Amsterdam, (2008)

         and to be published.

**Figure captions**

Figure 1.    $R(T)$ curves in a semi-log scale, for a series of films deposited on KTO substrates, the first one  being obtained for a sample cooled under gas process (GP), and the following ones for in-situ annealed samples at increasing temperatures $T_A$ for 10 minutes in vacuum ($2x10^{-6}$ Torr). The vertical arrows indicate the position of the minimum exhibited by $R(T)$ curves obtained for $T_A>500$ °C.

Figure 2.    Typical XRD $\theta$-$2\theta$ scan of a $Sr_{0.88}La_{0.12}CuO_2$ thin film deposited on a KTO substrate. The $y$-Log scale is used to magnify the background in order to show that the deposited film is single phase. In insert, $\Phi$-scan showing that the film is epitaxially grown.

Figure 3.    Lattice parameter $c$ versus annealing temperature $T_A$. In insert, normalized intensity of  XRD reflection (002) of the films versus angle $2\theta$ for no annealing (GP) and increasing annealing  temperature $T_A$. The full and empty symbols correspond to insulating and superconducting samples respectively.

Figure 4.    (a)  Resistive transition   for three $Sr_{0.88}La_{0.12}CuO_2$ thin films on KTO substrates, cooled down to different temperatures $T_A$ under Ar atmosphere and further annealed under vacuum at these  temperatures $T_A$  for 10mn. In insert, decrease of lattice parameter $c$ with increasing $T_A$. (b) $R(T)$ curve for a $Sr_{0.88}La_{0.12}CuO_2$ thin film on KTO substrate annealed at 615°C under vacuum. In insert, magnetic transition measured by $ac$ susceptibility (f=87hz, h=0.01Oe).

Figure 5.    $R(T)$ curves obtained for a series of films deposited on DSO substrates and annealed in vacuum at increasing temperatures. In insert, typical rocking curve of a film deposited on DSO.

Figure 6.    Variation versus annealing temperature $T_A$ , for the SLCO films deposited on DSO substrate, of parameter $c$ (left scale, filled symbols) and of the ratio, $I_{lc}/I_{lL}$ (right scale, open symbols), of the intensity of (002) reflections  of the  "long-c" phase and of the IL phase respectively. The three different symbols correspond to three different in-situ annealing times $t_A$. (b) Onset ($T_c$ *on* ) and offset ($T_c$ *off*) temperatures of the superconducting transition for the same films as in (a), as a function of the annealing temperature $T_A$ (see text).

Figure 7.   Partial XRD scan around (002) Bragg reflection of a $Sr_{0.88}La_{0.12}CuO_2$ IL thin film on DSO substrate a) without or b) with impurity "long c" phase. Superimposed are calculated curves.

Figure 8.   (a), (b).  R(T) curves of a single film deposited on a DSO substrate, submitted to a series of ex-situ annealing treatments, and measured after each treatment.

Figure 9.   (a) Hopping behaviour of the conduction of a $Sr_{0.88}La_{0.12}CuO_2$ thin film on DSO substrate, cooled under gas process, as shown by plotting $R_{in}$ (T) (displayed in fig. 8(a)) on a logarithmic scale  versus $T^{-\alpha}$. The solid line is the best fit of the data  to a function of the form $Aexp(T_0/T)^{\alpha}$  ($\alpha$= 0.18 , $T_0$=32 K).(b) Logarithmic low temperature dependence of $R(T)$ after the second and third  annealing treatment with onset of superconductivity.

Figure 10.   Variation as a function of the conductance at 300K of parameter $c$ (right scale)  and of the temperatures $T_{on}$  and $T_c(R=0)$ where $R$ is maximum or equal to zero respectively.

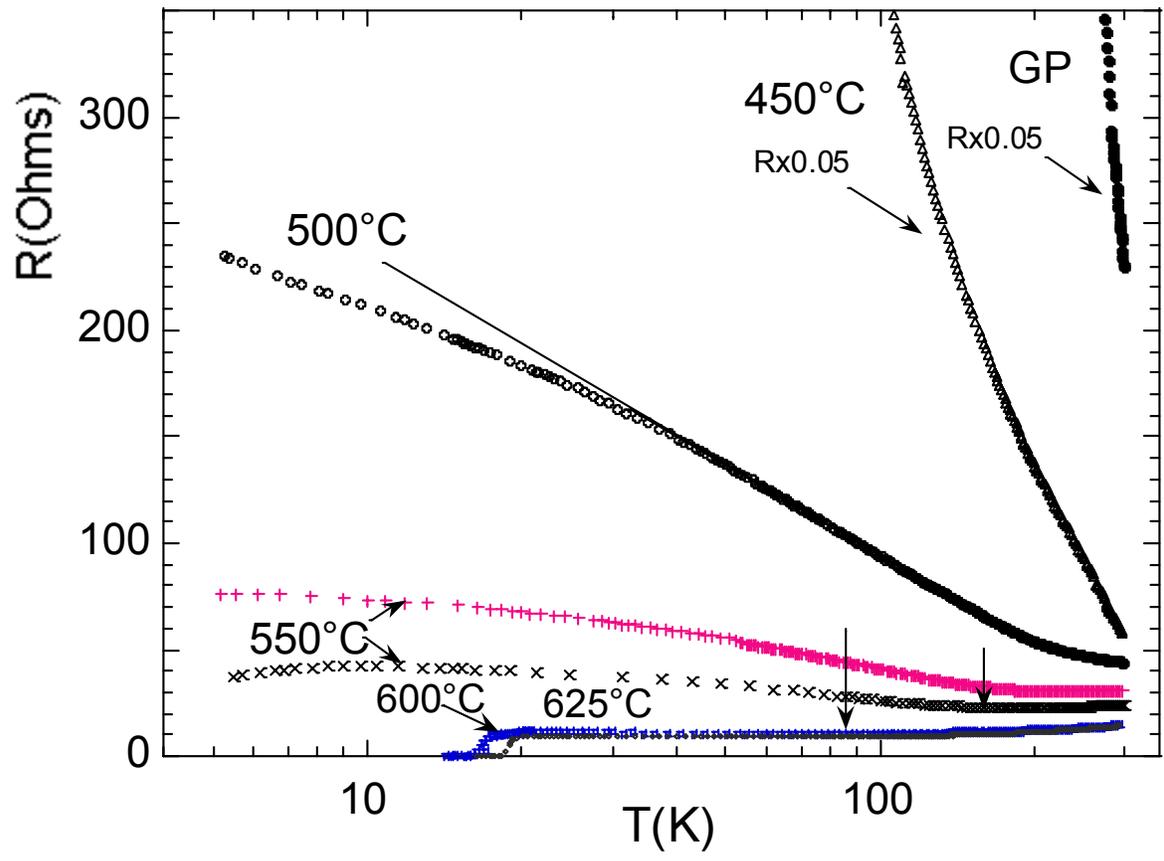

Fig.1. Z.Z. Li *et al.*

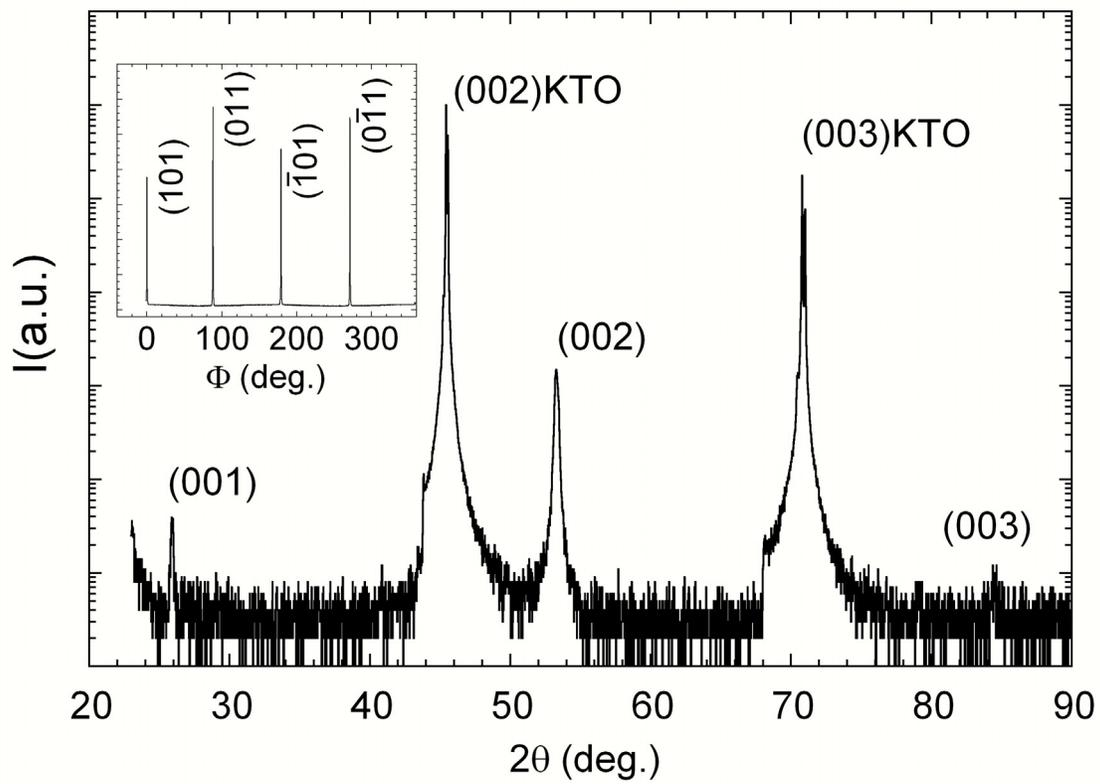

Fig. 2. Z.Z. Li *et al.*

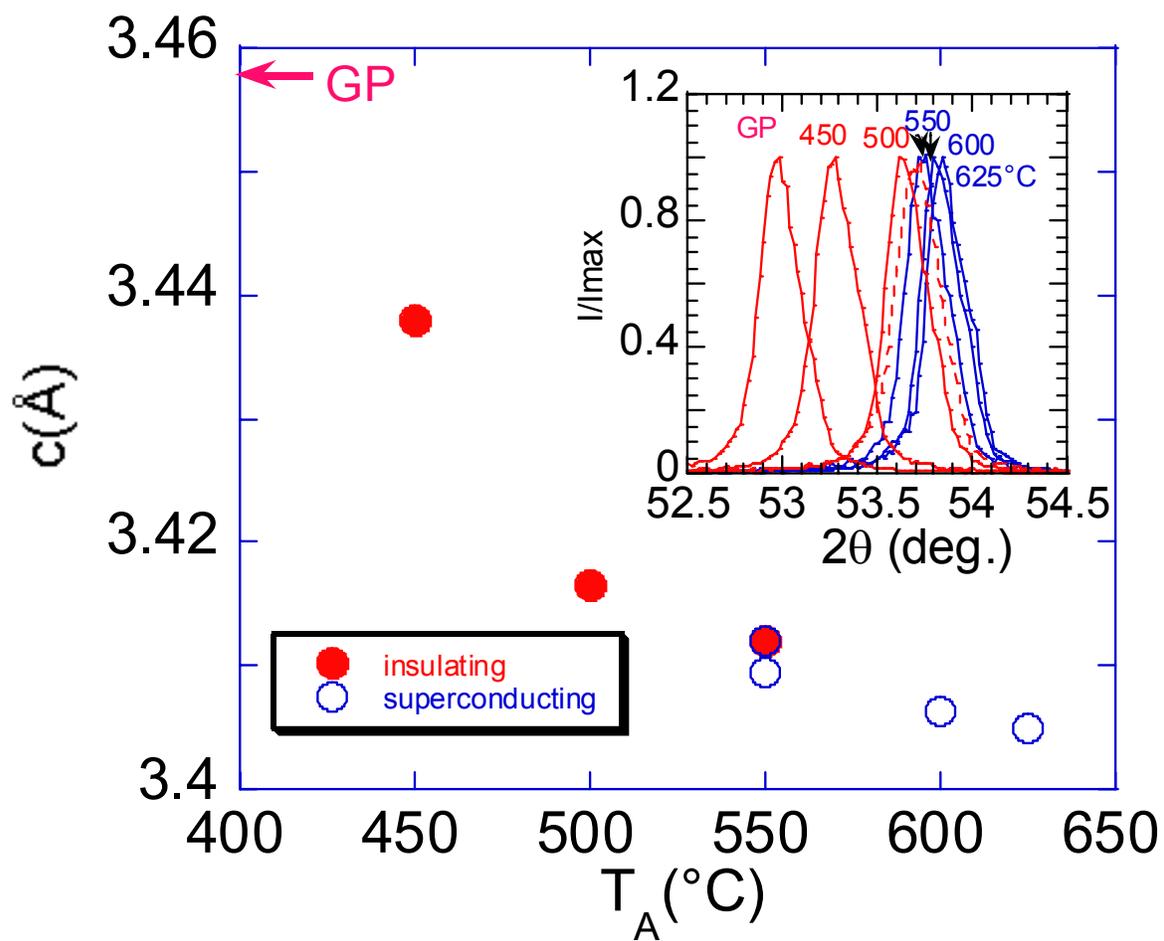

Fig. 3. Z.Z. Li *et al.*

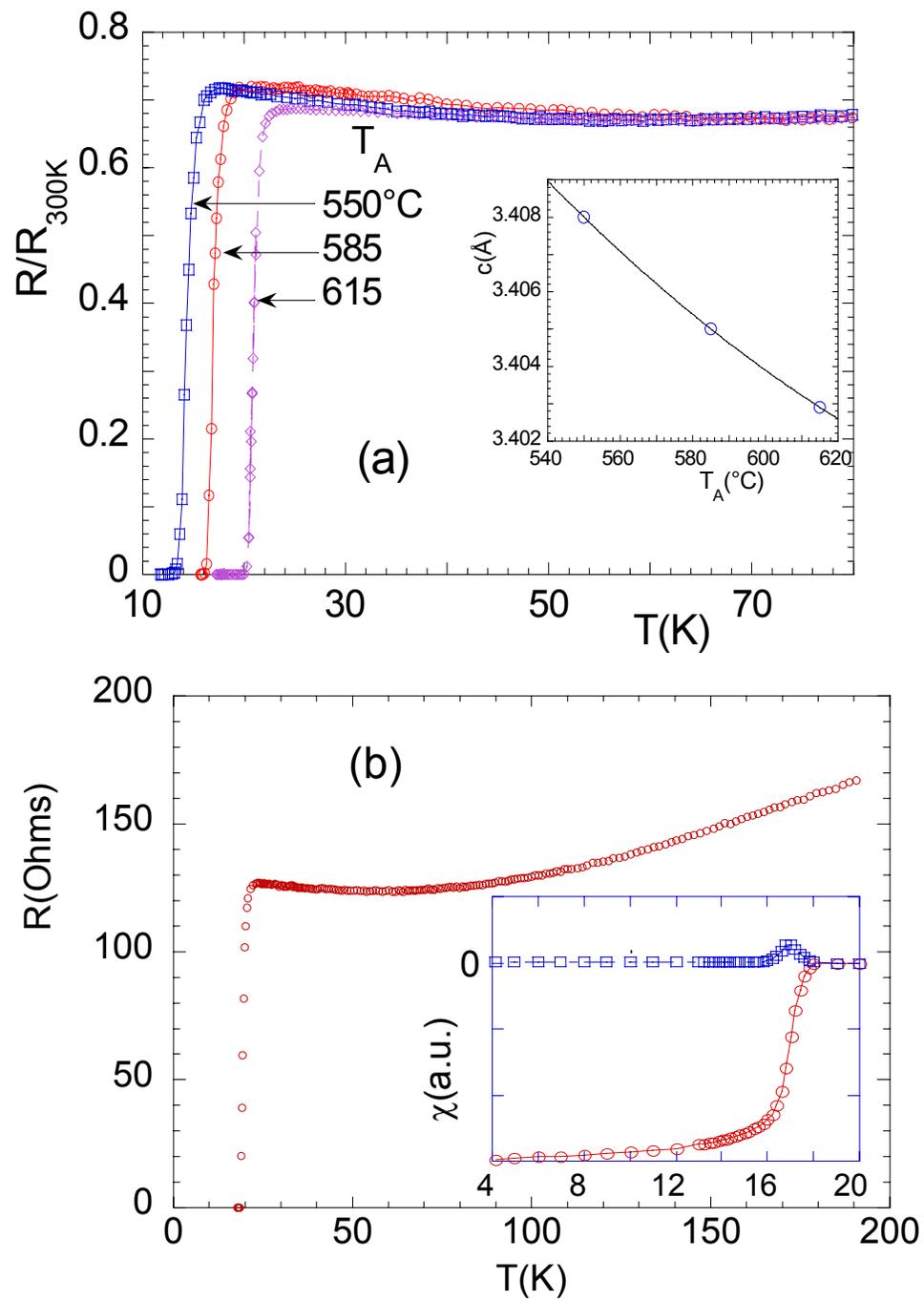

Fig.4. Z.Z. Li *et al.*

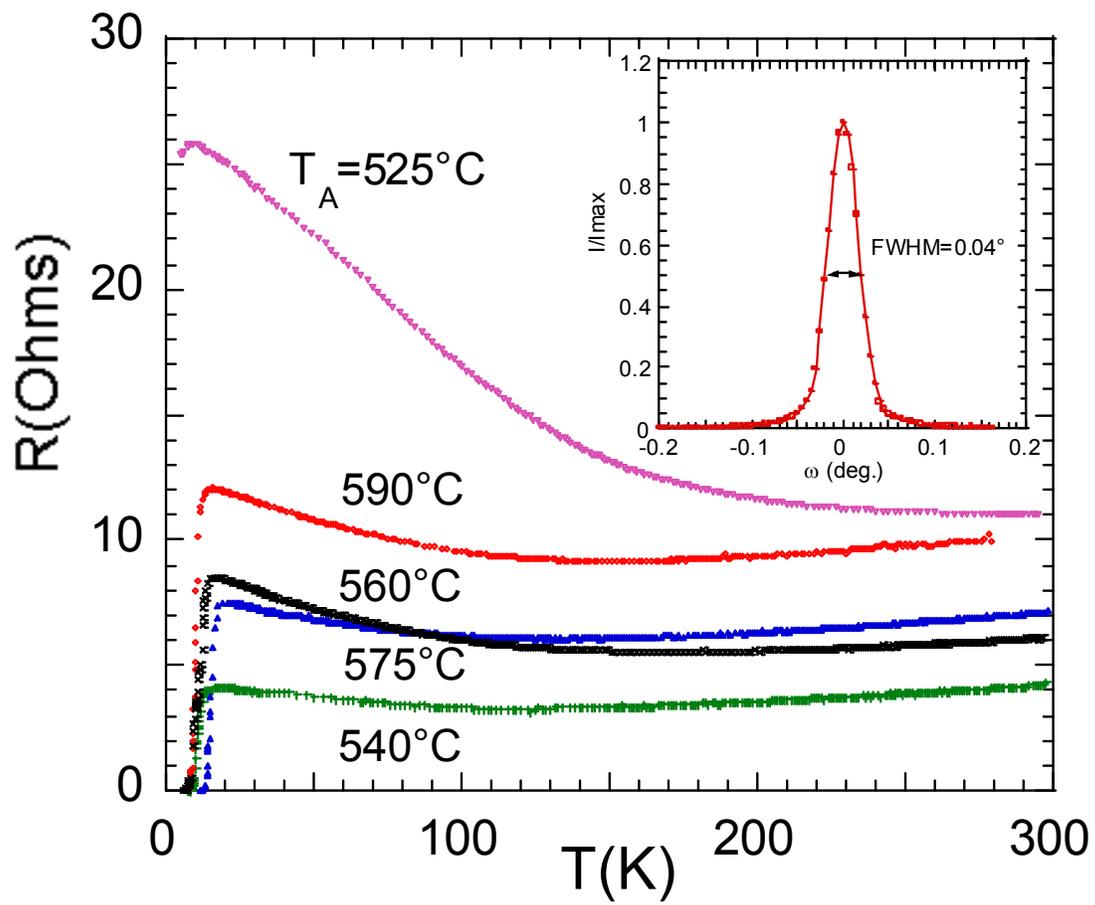

Fig. 5. ZZ Li *et al.*

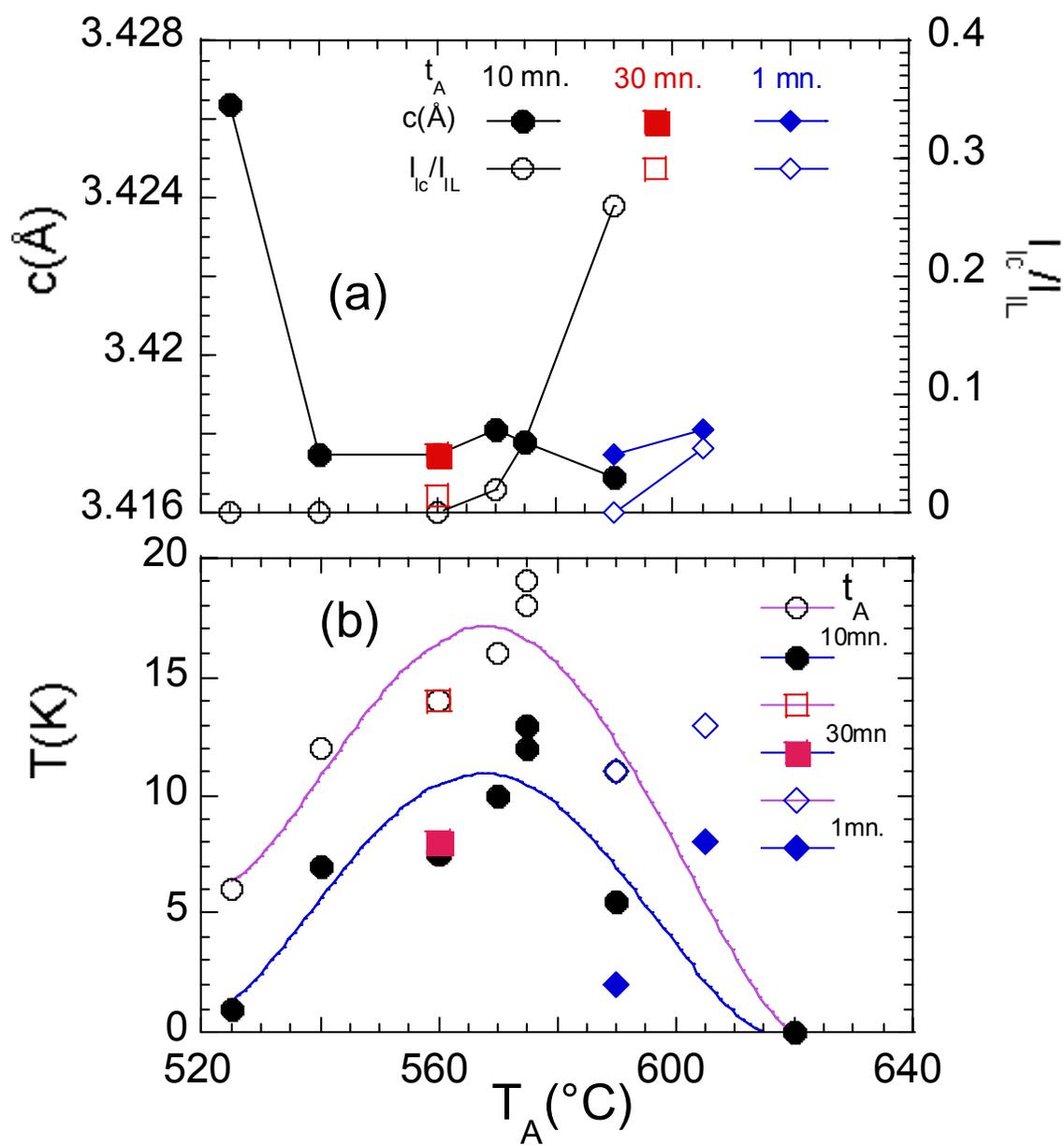

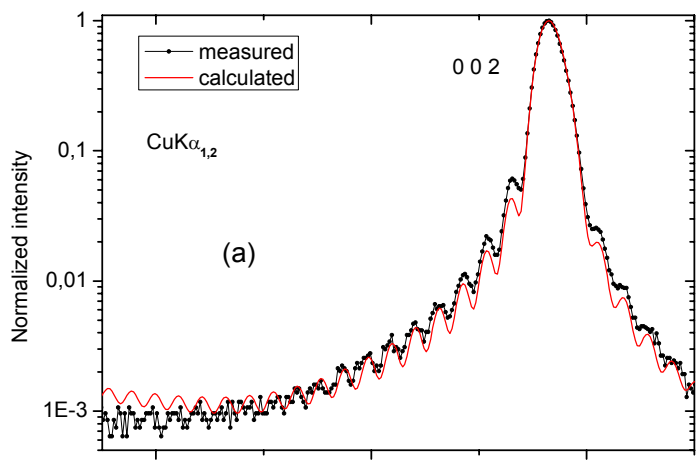

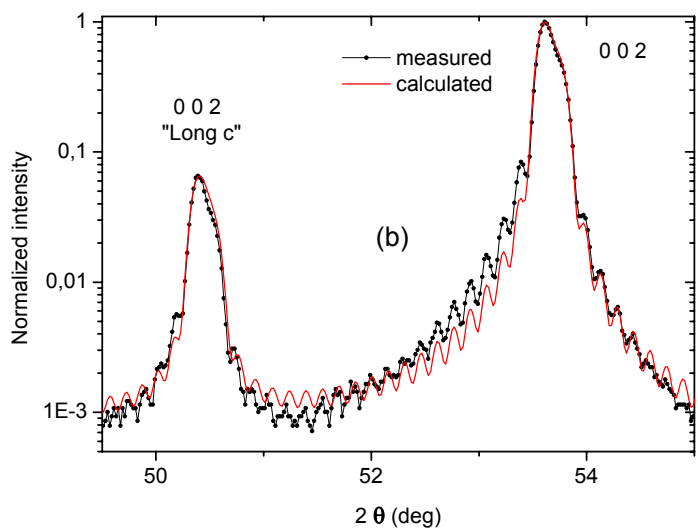

Fig. 7.  Z.Z. Li *et al.*

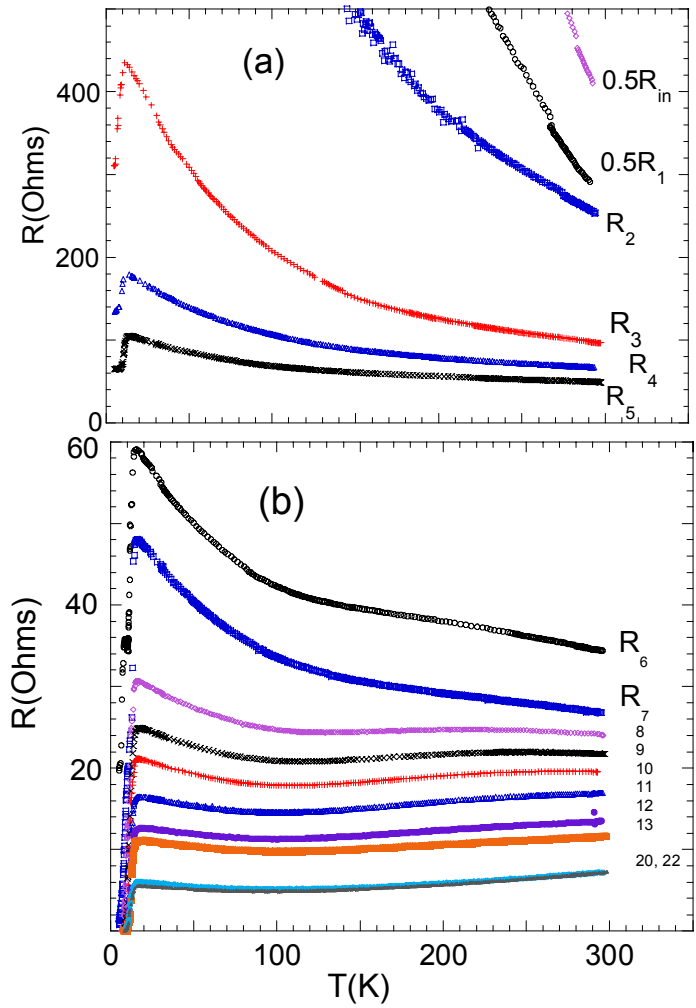

Fig. 8. Z.Z. Li *et al.*

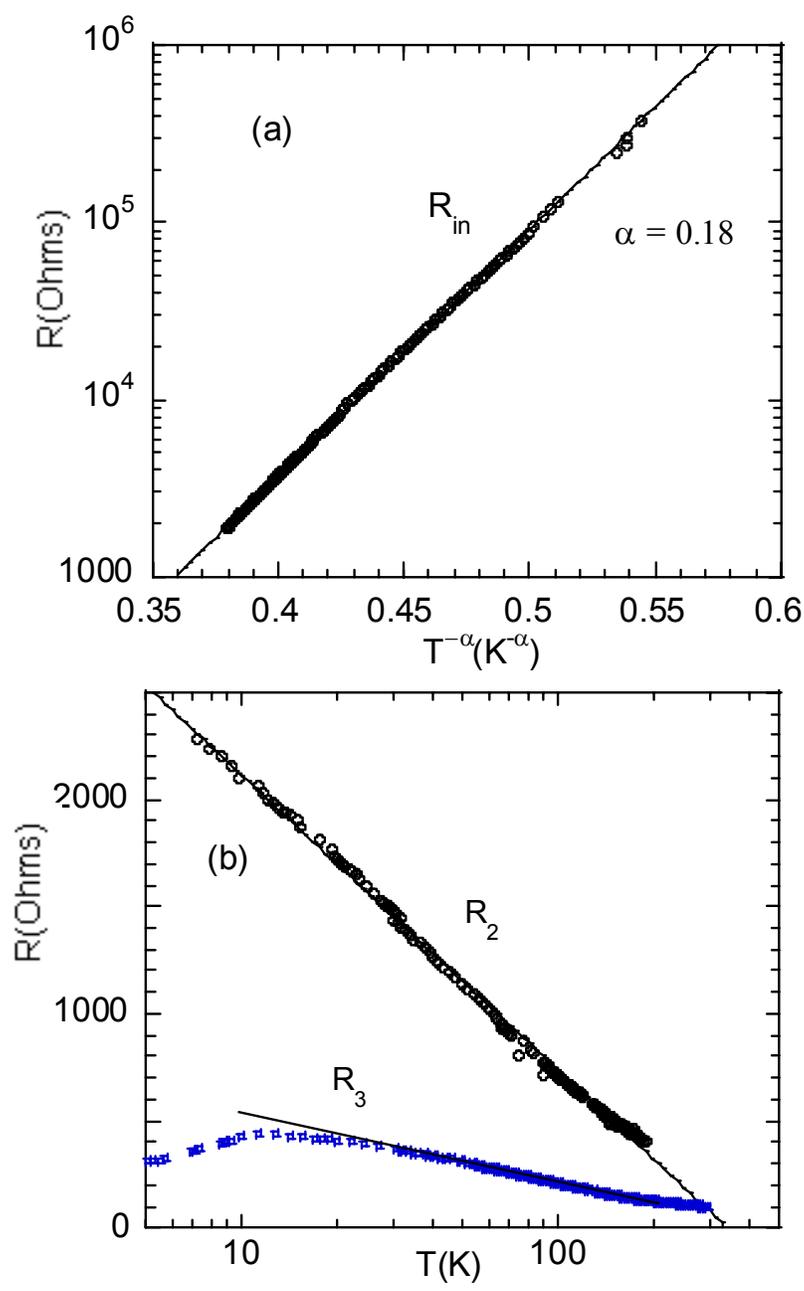

Fig. 9. Z.Z. Li *et al.*

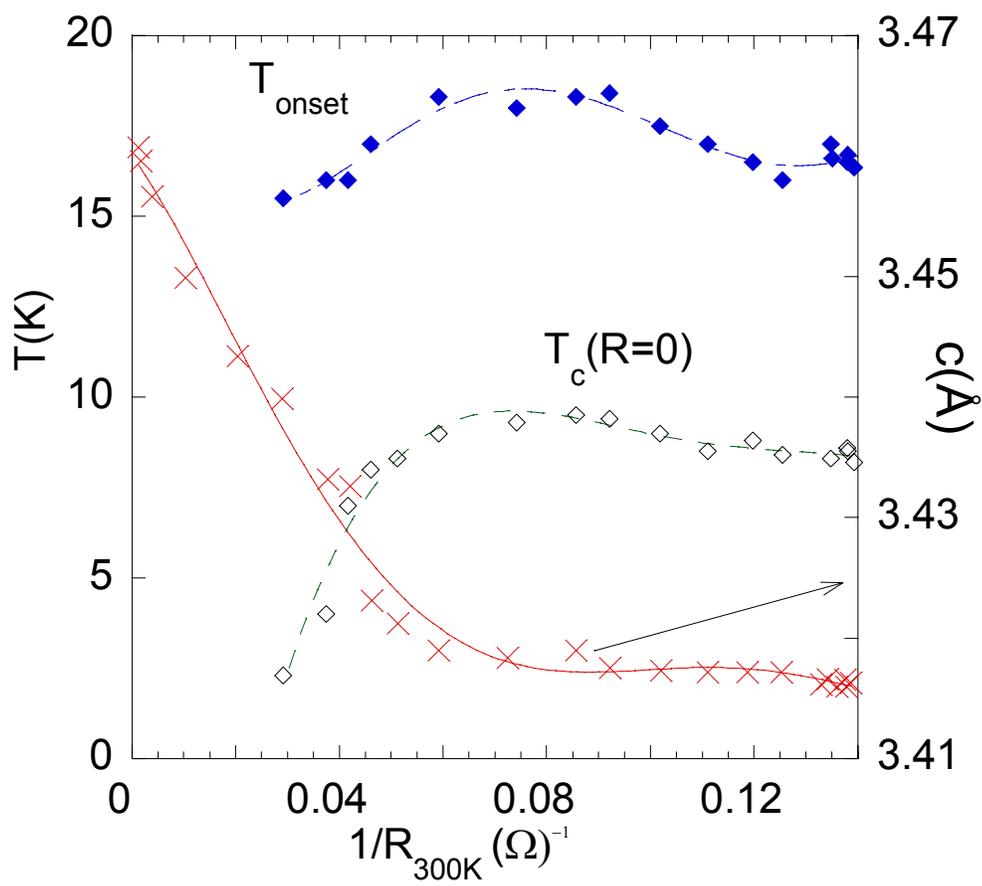

Fig. 10. Z.Z. Li *et al.*